\documentstyle[12pt]{article}
\newcommand{\be}{\begin{equation}}
\newcommand{\ee}{\end{equation}}

\begin{document}

\begin{titlepage}
\null
\begin{flushright}
KOBE-TH-96-02 \\
July 1996 \\
hep-th/9607176
\end{flushright}
\vspace{0.5cm}

\begin{center}
{\Large\bf Vacuum Wave Functional of Pure Yang-Mills \par}
{\Large\bf Theory and Dimensional Reduction\par}
\vspace{1.5cm}
\baselineskip=7mm

{\large Miyuki Kawamura\footnote{
E-mail address: myu@hetsun1.phys.kobe-u.ac.jp}
, Kayoko Maeda\footnote{
E-mail address: maeda@hetsun1.phys.kobe-u.ac.jp} \par}
{\sl 
  Graduate School of Science and Technology, 
  Kobe University \\
  Rokkodai, Nada, Kobe 657, Japan \par}
  
\vspace{0.7cm} 
{\large and} \\
\vspace{0.7cm}
 
{\large Makoto Sakamoto\footnote{
E-mail address: sakamoto@hetsun1.phys.kobe-u.ac.jp} \par}
{\sl 
  Department of Physics, Kobe University \\
  Rokkodai, Nada, Kobe 657, Japan \par}
\vspace{3.5cm}

{\large\bf Abstract}
\end{center}
\par

Working in a Hamiltonian formulation with $A_0 = 0$ gauge 
and also in a path integral formulation, we show that 
the vacuum wave functional of four-dimensional pure 
Yang-Mills theory has the form of the exponential of a 
{\it three}-dimensional Yang-Mills action. This result implies
that vacuum expectation values can be calculated in Yang-Mills
theory but one dimension lower. Our analysis reveals that this
dimensional reduction results from a stochastic nature of 
the theory. 

\end{titlepage}

%%%%%%%%%%%%%%%%%%%%%%%%%%%%%%%%%%%%%%%%%%%%%%%%%%%%%%%%%%%%%%%
\section{Introduction}

In spite of considerable success in describing high energy 
phenomena of QCD, the dynamics at low energies, such as 
confinement and chiral symmetry breaking, has not satisfactorily 
been understood yet. Nonperturbative approaches are essential 
to study the low energy dynamics because of the strong coupling 
nature. One of successful nonperturbative approaches is the 
lattice formulation. The strong coupling expansion on the lattice 
has succeeded to show an area law for Wilson loops\cite{Wilson}.
This strong coupling result seems to catch the essence of quark
confinement but a dissatisfaction of the lattice formulation 
is that the connection with continuum field theory has been 
established only numerically. Furthermore, the lattice 
formulation encounters a notorious fermion doubling 
problem\cite{doubling} and hence has much trouble in 
studying spontaneous chiral symmetry breaking. Thus, it will be 
important to develop nonperturbative methods in the continuum
formulation to study the low energy dynamics of nonabelian gauge
theories more transparently.

In this paper, we shall first examine the vacuum structure of 
pure Yang-Mills theory in the continuum Hamiltonian formulation,
which will be well suited to study the nonperturbative dynamics 
of the theory. Our aim is then to solve the ground state of the 
Schr\"odinger wave functional equation 
\be
H \Psi_0[A] = E_0 \Psi_0[A] ~ ~ ,
\label{1}
\ee
where $H$ is the Yang-Mills Hamiltonian. An approximate vacuum 
wave functional in the infrared regime has originally been 
suggested by Greensite\cite{Greensite} to be of the form
\be
\Psi_0[A] = ~N~  \exp \left\{ - \frac{\gamma}{g^4} 
            \displaystyle{\int d^3x (F^a_{ij})^2} \right\} ~~,
\label{2}
\ee
where $F^a_{ij}$ is a magnetic component of the field strength 
and $\gamma$ is a numerical constant. This result has been 
supported by other studies: A lattice version of the wave 
functional (\ref{2}) has been obtained in the strong coupling
expansion of the lattice Hamiltonian formulation\cite{LH} 
and also studied in the Monte Carlo method\cite{MC}. The wave
functional (\ref{2}) has been rederived in the continuum 
strong coupling expansion by Mansfield\cite{Mansfield}.
The form of the wave functional (\ref{2}) implies that vacuum 
expectation values can be calculated in a path integral 
representation of {\it three}-dimensional Yang-Mills theory.
Three-dimensional gauge theories have been studied by various 
authors. Polyakov\cite{Polyakov} has shown that three-dimensional
compact QED has a mass gap and confines electric charges. For
nonabelian gauge theories in three dimensions, 
Feynman\cite{Feynman} and recently many authors\cite{3YM} 
have discussed the existence of a mass gap. Greensite has shown 
an area law of Wilson loops in an analog gas approximation in his
original work\cite{Greensite}. The arguments of the 
$D=4 \rightarrow D=3$ dimensional reduction might work again for
resulting three-dimensional Yang-Mills theory and then the theory 
would reduce to two-dimensional Yang-Mills theory, which exhibits 
confinement trivially\cite{Halpern, Mansfield}. The vacuum wave
functional (\ref{2}) strongly suggests that in a strong coupling 
regime the vacuum consists of random magnetic fluxes, which has 
been discussed to be a necessary and sufficient condition for 
confinement\cite{random, Olesen}.

Although the wave functional (\ref{2}) is a good candidate for 
the vacuum and possesses desired properties, all of the previous 
works have not verified that it is really the vacuum, i.e., the
lowest energy state because they have looked for solutions of the
Schr\"odinger equation (\ref{1}) by taking appropriate ansatzs of
the vacuum wave functional. In this paper, we shall show that the
wave functional (\ref{2}) is the lowest energy state in a more 
convincing way. Our approach reveals that the $D=4 \rightarrow 
D=3$ dimensional reduction results from a stochastic nature of 
the theory: We shall show that four-dimensional Yang-Mills theory
in the infrared regime is equivalently described by the following 
stochastic system:
\be
\frac{\partial A^a_i(x,t)}{\partial t} = - \frac{g^2}{2} 
\left.
\frac{\delta S_{3YM}[A]}{\delta A^a_i(x)}
\right|_{A_i(x) = A_i(x,t)} 
+ \eta^a_i(x,t)  ~~,
\label{3}
\ee
where $\eta^a_i$ is a Gaussian white noise and $S_{3YM}$ is a 
three-dimensional Yang-Mills action. In the equilibrium limit 
$t \rightarrow \infty$, this system has been shown to be 
equivalent to the quantum theory with the action 
$S_{3YM}$\cite{Parisi-Wu}.

This paper is organized as follows: In Sec. 2, we solve a 
regularized version of the Schr\"odinger equation (\ref{1}) 
and show that the vacuum wave functional takes to be of the form 
(\ref{2}) in the limit of the cutoff $s \rightarrow 0$. 
In Sec. 3, Euclidean four-dimensional Yang-Mills theory can 
equivalently be described by the Langevin equation (\ref{3})
in the limit $s \rightarrow 0 $. Sec. 4 is devoted to conclusion.

%%%%%%%%%%%%%%%%%%%%%%%%%%%%%%%%%%%%%%%%%%%%%%%%%%%%%%%%%%%%%%%%%%

\section{Vacuum Wave Functional}

We shall consider pure Yang-Mills theory whose Lagrangian is given 
by 
\be
{\cal L} = - \frac{1}{4 g^2} F^a_{\mu\nu} F^{\mu\nu a} ~~,
\label{4}
\ee
where $F^a_{\mu\nu}$ is the field strength defined by 
\be
F^a_{\mu\nu} = 
\partial_{\mu} A^a_{\nu} -  \partial_{\nu} A^a_{\mu}
+ f^{abc} A^b_{\mu} A^c_{\nu}  ~~.
\label{5}
\ee
For our purposes, it is convenient to choose the $A_0 = 0$ gauge. 
In the Schr\"odinger representation, the (unregulated) 
Hamiltonian is then given by 
\be
H = \int d^3 x \left\{ -\frac{g^2}{2} 
\frac{\delta^2}{\delta A^a_i(x) \delta A^a_i(x)} 
+ \frac{1}{4 g^2} \left(F^a_{ij} (x) \right)^2 \right\} ~~.
\label{6}
\ee
Here Latin indices $i,j,k$ etc. run over the values 1,2, and 3.
In the $A_0 = 0$ gauge, the wave functional $\Psi[A]$ has to be 
subject to the Gauss' law constraint
\be
\left(D_i \frac{\delta}{\delta A_i(x)} \right)^a \Psi[A] = 0 ~~,
\label{7}
\ee
where $D_i$ denotes a covariant derivative. This constraint 
simply means that the wave functional is invariant under 
time-independent gauge transformations. The Schr\"odinger 
equation (\ref{1}) with the Hamiltonian (\ref{6}) needs 
regularization because it contains a product of two functional 
derivatives at the same spatial point. 
\be
\Delta \equiv \int d^3 x 
\frac{\delta^2}{\delta A^a_i(x) \delta A^a_i(x)}  ~~.
\label{8}
\ee
To make the differential operator (\ref{8}) well defined, we
replace $\Delta$ by the following differential 
operator\cite{Mansfield}:
\be
\Delta(s) \equiv \int d^3 x d^3 y 
\frac{\delta}{\delta A^a_i(x)} K^{ab}_{ij} (x,y;s) 
\frac{\delta}{\delta A^b_j(y)}  ~~.
\label{9}
\ee
The kernel $K^{ab}_{ij} (x,y;s)$ is required to satisfy a heat
equation 
\be
\frac{\partial}{\partial s} K^{ab}_{ij} (x,y;s) \\
= \left[ \delta_{ik} \left(D^2(x) \right)^{ac} 
       - \left(D_i(x) D_k(x) \right)^{ac} 
       - 2 f^{acd} F^d_{ik}(x) \right] K^{cb}_{kj} (x,y;s)  ~~,
\label{10} 
\ee
with the initial condition 
\be
\displaystyle{\lim_{s \rightarrow 0}}~ K^{ab}_{ij} (x,y;s)  = 
~\delta_{ij}~ \delta^{ab}~ \delta^3(x-y) ~~.
\label{11}
\ee
Taking $s$ small but nonzero in Eq.(\ref{9}) gives a regularized
operator of $\Delta$. ( In the naive limit $ s \rightarrow 0 ,
\Delta(s)$ is reduced to $\Delta$. ) It should be emphasized 
that the regularized operator $\Delta(s)$ preserves gauge 
invariance and (three-dimensional) Lorentz invariance. 

The heat equation (\ref{10}) can be solved by the standard 
technique\cite{heat kernel}. Acting $\Delta(s)$ on 
three-dimensional integrals of local functions will give an 
expansion in powers of $s$ and may contain inverse powers of $s$,
which diverge as $s \rightarrow 0 $. These powers of $s$ may be 
determined from dimensional analysis and gauge invariance. We
have, for example, 
\begin{eqnarray}
&&\displaystyle{ 
\int d^3x d^3y 
\frac{\delta}{\delta A^a_i(x)} K^{ab}_{ij} (x,y;s) 
\frac{\delta}{\delta A^b_j(y)} \int d^3z (F^c_{kl}(z))^2 }
         \nonumber\\
&&= \displaystyle{
\int d^3 x \left\{ 
  \frac{\alpha_1}{s^{5/2}} 
+ \frac{\alpha_2}{s^{1/2}} (F^a_{ij})^2 \right. }   \nonumber \\
&& \hspace{1.5cm}
  \left.+ s^{1/2} \left( \alpha_3(D^{ab}_i F^b_{ij})^2 
          + \alpha_4 f^{abc} F^a_{ij}F^b_{jk}F^c_{ki} \right)
          + {\cal O}(s^{3/2}) \right\}   ~~ ,
\label{12}
\end{eqnarray}
where $\alpha_n$'s are numerical constants. The first two 
coefficients are given by 
\be
\alpha_1 = \frac{3 \dim G}{2 \pi^{3/2}}  ~~, \\ ~~~~~~~
\alpha_2 = - \frac{11 C_2(G)}{24 \pi^{3/2}}  ~~,
\label{13}
\ee
where $\dim G$ is the number of generators of the gauge group $G$
and $C_2(G)$ is given by $f^{acd} f^{bcd} = C_2(G) \delta^{ab}$.

We now have a regularized Hamiltonian 
\be
H[A;s]  \\
= \int d^3 x \left\{
 - \frac{g^2}{2} \int d^3 y 
   \frac{\delta}{\delta A^a_i(x)} K^{ab}_{ij} (x,y;s) 
   \frac{\delta}{\delta A^b_j(y)}
+ \frac{1}{4 g^2} \left(F^a_{ij} (x) \right)^2  \right\}   ~~.
\label{14}
\ee
Let us rewrite the regularized Hamiltonian (\ref{14}) 
into the form  
\be
H[A;s]  \\
= \int d^3 x d^3 y ~
   {Q^a_i}^{\dagger} (x) K^{ab}_{ij} (x,y;s)  Q^b_j (y)
 + \Gamma[A;s] ~~.                                    
\label{15}
\ee
The operators $Q^a_i$ and ${Q^a_i}^{\dagger}$ are defined by 
\begin{eqnarray}
Q^a_i(x) = i \frac{g}{\sqrt{2}} \left(
           \frac{\delta}{\delta A^a_i(x)} 
         + \frac{1}{2} 
           \frac{\delta S_{3YM}[A]}{\delta A^a_i(x)} \right) ~~, 
\nonumber \\
{Q^a_i}^{\dagger}(x) = i \frac{g}{\sqrt{2}} \left(
           \frac{\delta}{\delta A^a_i(x)} 
         - \frac{1}{2} 
           \frac{\delta S_{3YM}[A]}{\delta A^a_i(x)} \right) ~~,
\label{16}
\end{eqnarray}
where 
\be
S_{3YM}[A] = \frac{24 \pi^{3/2} s^{1/2}}{11 C_2(G) g^4}
             \int d^3 x \left(F^a_{ij} (x) \right)^2   ~~.
\label{17}
\ee
A key observation is that $\Gamma[A;s]$ in Eq.(\ref{15}) 
vanishes in the naive limit $s \rightarrow 0$. It is easy to see
from the formula (\ref{12}) that $\Gamma[A;s]$ has the form 
\be
\Gamma[A;s] = \int d^3 x \left\{ 
s ( \beta_1 (D^{ab}_i F^b_{ij})^2 
  + \beta_2 f^{abc} F^a_{ij}F^b_{jk}F^c_{ki} )
  + {\cal O}(s^2) \right\} ~~,
\label{18}
\ee
up to a field independent constant. The $\Gamma[A;s]$ contains
only higher dimensional terms with positive powers of $s$. Thus,
in the naive limit $s \rightarrow 0$, $\Gamma[A;s]$ 
vanishes\footnote{ In taking the limit $s \rightarrow 0$, we
have to take the $s$ dependence of the gauge coupling $g$ into
account because $g$ in the Hamiltonian (\ref{14}) is a bare 
coupling and should depend on the cutoff $s$ for the theory to 
be renormalizable. We can show that $\Gamma[A;s]$ still 
vanishes as $s \rightarrow 0$ even if the $s$ dependence of $g$
is taken into account.} (up to an irrelevant constant).
Therefore, in the limit $s \rightarrow 0$, the Hamiltonian 
(\ref{14}) may be replaced by 
\be
\overline{H}[A;s] \equiv \int d^3 x d^3 y ~
 {Q^a_i}^{\dagger} (x) K^{ab}_{ij} (x,y;s)  Q^b_j (y) ~~,
\label{19}
\ee
up to an irrelevant constant. Since the kernel $K^{ab}_{ij}$ is 
positive definite, $\overline{H}$ is positive semi-definite. 
Thus, a zero energy eigenstate of $\overline{H}$, 
if any, is the lowest 
energy state, i.e., the vacuum\footnote{
Greensite\cite{Greensite} has found a zero energy solution 
to the {\it unregulated} 
Hamiltonian. The wave functional is not, however, normalizable
and hence does not seem to have physical meaning.}, 
\be
\overline{H} \Psi_0[A] = 0 ~~.
\label{20}
\ee
It follows from the form (\ref{19}) that the above equation is
equivalent to solve 
\be
Q^a_i (x) \Psi_0[A] = 0  ~~, 
\label{21}
\ee
which leads to a solution
\be
\Psi_0[A] = N~  
            \exp \left\{ -\frac{1}{2} S_{3YM}[A] \right\}  ~~,
\label{22}
\ee
as announced in the introduction. It should be emphasized that
we have not assumed any specific form of the vacuum wave 
functional to derive Eq.(\ref{22}). Since $S_{3YM}$ in 
Eq.(\ref{17}) is positive semi-definite, the wave functional 
$\Psi_0[A]$ is normalizable, as it should be.

Let $F[A]$ be any functional of $A^a_i$. The vacuum expectation 
value of $F[A]$ can be expressed as 
\be 
\int {\cal D}A_i~ \Psi_0^{\ast}[A] F[A] \Psi_0[A] \\
= N^2 \int {\cal D}A_i ~ F[A]~ 
 \exp \left\{- S_{3YM}[A] \right\}  ~~.
\label{23}
\ee
The last expression is identical to a path integral representation 
of three-dimensional Yang-Mills theory. Vacuum expectation values
of any physical operators have to be independent of the cutoff 
$s$, so that the coupling constant $g$ should be regarded as a 
function of $s$\footnote{ In our field definition (\ref{5}), 
there is no wave function renormalization.}. It follows from 
Eqs.(\ref{17}) and (\ref{23}) that the $s$ dependence of $g$ 
should be given by\cite{Greensite, Mansfield}
\be
{g(s)}^4 s^{-1/2} = s ~{\rm - ~independent} ~~. 
\label{24}
\ee

%%%%%%%%%%%%%%%%%%%%%%%%%%%%%%%%%%%%%%%%%%%%%%%%%%%%%%%%%%%%%%%%%%

\section{Stochastic Quantization Point of View}

In the previous section, we have derived the vacuum wave 
functional (\ref{22}) in the Hamiltonian formulation. In what 
follows, we shall show that the same conclusion (\ref{22}) 
can be obtained from a stochastic quantization point of 
view\footnote{For reviews, see Ref. \cite{stochastic}.}.

Let us start with the following Langevin equation:
\be
\frac{\partial A^a_i(x,t)}{\partial t} 
= - \frac{g^2}{2} 
  \left.
  \frac{\delta S_{3YM}[A]}{\delta A^a_i(x)} 
  \right|_{A_i(x) = A_i(x,t)}
  + \eta^a_i(x,t)  ~~, 
\label{25}
\ee
where $\eta^a_i$ is a Gaussian white noise and $S_{3YM}[A]$ is 
given in Eq.(\ref{17}). 
The average over $\eta^a_i$ is defined by 
\be
\left< F[A^{\eta}] \right>_{\eta} 
= N' \int {\cal D} \eta_i~ F[A^{\eta}]~ 
  \displaystyle{ 
    \exp \left\{
       - \frac{1}{2 g^2} \int d^3x dt 
       (\eta^a_i(x,t))^2 \right\} }  ~~,
\label{26}
\ee
where $F$ is an arbitrary function of $A^a_i$, $N'$ is a 
normalization constant, and $A^\eta$ exhibits the $\eta$ 
dependence as a solution of the Langevin equation (\ref{25}).
We shall now show that the $\eta$ average (\ref{26}) can be 
rewritten as 
\be
\left< F[A^{\eta}] \right>_{\eta} 
= N' \int {\cal D} A_\mu  ~ F[A] ~ \delta(A_0)~
  \exp \left\{- S_{4YM}[A] \right\}  ~~,
\label{28}
\ee
where $S_{4YM}$ is the (Euclidean) four-dimensional Yang-Mills 
action. The right hand side of Eq.(\ref{28}) is nothing but a 
path integral representation of Euclidean four-dimensional 
Yang-Mills theory in the $A_0 = 0$ gauge. The equality in 
Eq.(\ref{28}) should be understood in the same sense that $H$ in 
Eq.(\ref{14}) is replaced by $\overline{H}$ in Eq.(\ref{19}) in 
the limit $s \rightarrow 0$. To show the relation (\ref{28}), 
we will change the variables from $\eta^a_i$ to $A^a_i$ in 
Eq.(\ref{26}) through the equation (\ref{25}). Then, the 
exponent of Eq.(\ref{26}) can be rewritten as\footnote{ 
In Eqs. (\ref{30}), (\ref{32}) and (\ref{33}), we have not
taken account of the $s$ dependence of the gauge coupling $g$ 
given in Eq. (\ref{24}). Even if the $s$ dependence of $g$ has
been taken account of, the leading terms shown in
Eqs. (\ref{30}), (\ref{32}) and (\ref{33}) are still correct.}   
\begin{eqnarray}
- \frac{1}{2 g^2} \int d^3x dt (\eta^a_i(x,t))^2 
&=& - \frac{1}{2 g^2} \int d^3x dt~ 
      \left(    
           \frac{\partial A^a_i(x,t)}{\partial t} 
         + \frac{g^2}{2} \left.
           \frac{\delta S_{3YM}[A]}
                {\delta A^a_i(x)}  \right|_{A_i(x) = A_i(x,t)} 
      \right)^2 
                \nonumber\\
&=& - \frac{1}{2 g^2} \int d^3x dt~ 
      \left(\frac{\partial A^a_i(x,t)}{\partial t}\right)^2 
  + {\cal O}(s)  ~~,
\label{30}
\end{eqnarray}
where we have dropped a total derivative term in the last 
equality. We next calculate the Jacobian. 
\begin{eqnarray}
\det \left(\frac{\delta \eta^a_i}{\delta A^b_j}\right) 
&=&  \det \left( \frac{\partial}{\partial t} + \frac{g^2}{2}
   \frac{\delta^2 S_{3YM}[A]}{\delta A^a_i \delta A^b_j} \right) 
  \nonumber\\
&=& \exp  \left\{  \int d^3 x dt~ \frac{g^2}{4}
     \left.
     \frac{\delta^2 S_{3YM}[A]}
          {\delta A^a_i(x) \delta A^a_i(x)}
          \right|_{A_i(x) = A_i(x,t)} 
     \right\} ~~,
\label{31}
\end{eqnarray}
where we have chosen the retarded Green's function of 
$\frac{\partial}{\partial t}$ to show the last equality in 
Eq.(\ref{31})\cite{stochastic}, and omitted a field independent 
constant in Eq.(\ref{31}). The expression on the right hand
side of Eq.(\ref{31}) is, however, ill defined because it 
contains a product of two functional derivatives at the same 
spatial point, as found in the Hamiltonian (\ref{6}). 
According to the prescription discussed in the previous section, 
we will regularize the product of two functional derivatives.
A regularized Jacobian is then given by 
\begin{eqnarray}
\det  \left(\frac{\delta \eta^a_i}{\delta A^b_j}\right)_{reg} 
&=& \exp \left\{ \int d^3x d^3y dt~ 
     \frac{g^2}{4} \left.
     \frac{\delta}{\delta A^a_i(x)} K^{ab}_{ij} (x,y;s) 
     \frac{\delta}{\delta A^b_j(y)} 
     S_{3YM}[A]\right|_{A_i(x) = A_i(x,t)} \right\}  \nonumber\\
&=& \exp  \left\{ - \int d^3x dt~ 
    \frac{1}{4 g^2} 
    \left(F^a_{ij}(x,t)\right)^2 
    + {\cal O}(s) \right\} ~~, 
\label{32}
\end{eqnarray}
where we have used the formula (\ref{12}) and ignored an 
irrelevant constant. Combining Eqs. (\ref{30}) and (\ref{32}),
we finally arrive at the conclusion (\ref{28}), i.e.,
\begin{eqnarray}
\left< F[A^\eta] \right>_\eta  
&=& N' \int {\cal D}A_i ~F[A]~ 
 \det \left(\frac{\delta \eta}{\delta A}\right)_{reg}
 \exp \left\{ \displaystyle{
    - \frac{1}{2 g^2} 
    \int d^3x dt \left( 
    \dot{A}^a_i 
  + \frac{g^2}{2} \frac{\delta S_{3YM}}{\delta A^a_i}
                 \right)^2 } \right\}
   \nonumber\\
&\propto& \int {\cal D}A_i~ F[A]~ 
 \exp \left\{-  \displaystyle{
      \int d^3x dt \left\{ \frac{1}{2 g^2} (\dot{A}^a_i)^2
    + \frac{1}{4 g^2} (F^a_{ij} )^2 
    + {\cal O}(s) \right\} } \right\} 
   \nonumber\\
&=& \int {\cal D}A_\mu ~F[A]~ \delta(A_0)~ 
 \exp \left\{- \displaystyle{ \int d^3x dt \left\{ 
   \frac{1}{4 g^2} (F^a_{\mu\nu} )^2 
   + {\cal O}(s) \right\}} \right\} ~~.
\label{33}
\end{eqnarray}
Thus, we may conclude that four-dimensional Yang-Mills theory can
equivalently described by the stochastic system governed by the 
Langevin equation (\ref{25}). As seen above, $t$ in the Langevin 
equation (\ref{25}) corresponds to the Euclidean time, though
$t$ is usually regarded as a fictitious time in a stochastic 
quantization point of view\cite{stochastic}.

In the operator language, the right hand side of Eq.(\ref{28})
may be written as 
\be
N' \int_{\hspace{-2.3mm}~_{A_\mu(T) = A_\mu(0)}} 
   \hspace{-8mm} {\cal D}A_\mu ~F[A] ~\delta(A_0)~ 
   \exp \left\{- \displaystyle{
           \int d^3x \int^T_0 dt 
           \frac{1}{4 g^2} (F^a_{\mu\nu})^2 } \right\}
= {\rm Tr} \left( F[A] {\bf e}^{-T \overline{H}} \right) ~~,
\label{34}
\ee
where we have chosen a periodic boundary condition, 
$A^a_\mu(T) = A^a_\mu(0)$. Taking the limit $T \rightarrow \infty$ 
gives
\begin{eqnarray}
\lim_{T \rightarrow \infty} 
{\rm Tr} \left( F[A] {\bf e}^{-T \overline{H}} \right)  
&=& \lim_{T \rightarrow \infty} \sum_{n} {\bf e}^{-T E_n}
    \left<n| F[A] |n \right> \nonumber \\
&\simeq& \lim_{T \rightarrow \infty} {\bf e}^{-T E_0}
    \left<0| F[A] |0 \right> \nonumber \\
&=& \int {\cal D}A_i ~\Psi^{\ast}_0[A] F[A] \Psi_0[A]  ~~.
\label{35}
\end{eqnarray}
On the other hand, the Parisi-Wu dimensional reduction implies
that the left hand side of Eq.(\ref{28}) will be reduced 
to\cite{Parisi-Wu} 
\begin{eqnarray} 
\lim_{T \rightarrow \infty} \left< F[A^\eta] \right>_\eta
&=& \lim_{T \rightarrow \infty} N' \int {\cal D}\eta_i~ F[A^\eta]~
   \exp \left\{ \displaystyle{ - \frac{1}{2 g^2} 
             \int d^3x \int^T_0 dt (\eta^a_i)^2 }
             \right\} \nonumber\\
&=& N' \int {\cal D}A_i~ F[A] ~ 
    \exp \left\{- S_{3YM}[A] \right\}  ~~.
\label{36}
\end{eqnarray}
Comparing Eq.(\ref{35}) with Eq.(\ref{36}), we arrive at the 
same vacuum wave functional (\ref{22}) that we have derived in 
the Hamiltonian formulation. We therefore conclude that the 
vacuum wave functional (\ref{22}) results from a stochastic 
nature of the theory. It is interesting to note that if we 
regard the Langevin equation (\ref{25}) as a mapping of 
$A^a_i$ to $\eta^a_i$, it is a kind of Nicolai 
mapping\footnote{ 
Claudson and Halpern\cite{CH} have given different Nicolai 
maps for Yang-Mills theory in four dimensions,
based on the Chern-Simons action, which have been
explicitly checked to all orders by Bern and Chan\cite{BernChan}.
Connections with our results are unclear.}\cite{Nicolai}, 
which implies the existence of a hidden 
supersymmetry\cite{SUSY}.

%%%%%%%%%%%%%%%%%%%%%%%%%%%%%%%%%%%%%%%%%%%%%%%%%%%%%%%%%%%%%%%%%

\section{Conclusion}

We have shown that the vacuum wave functional has the form 
(\ref{22}) in the naive limit $s \rightarrow 0$. This does not, 
however, mean that the wave functional (\ref{22}) is an exact
expression for the vacuum, as discussed in Ref.\cite{Greensite}.
We have dropped higher dimensional terms because they are 
proportional to positive powers of $s$ and hence might vanish in 
the limit $s \rightarrow 0$. Taking the naive limit 
$s \rightarrow 0$ can, however, be dangerous because the scaling
behavior in Eq.(\ref{24}) is different from what one expects in
the weak coupling regime. This situation seems to be similar to 
what one finds in the strong coupling expansion of the lattice
gauge theory\cite{Wilson}. Although we believe that our results 
are qualitatively correct, it is important to show how our results
connect with the weak coupling regime of the theory to make our 
analysis quantitative. 

Finally we would like to make a comment on dimensional reduction.
A simple picture of confinement has previously been 
proposed\cite{reduction, Olesen}: Random magnetic fluxes are 
dominating field configurations in a confining QCD vacuum and 
the theory exhibits the Parisi-Sourlas dimensional reduction of 
the type $D=4 \rightarrow D=2$\cite{Parisi-Sourlas} in the 
infrared regime. It is well known that two-dimensional QCD 
trivially confines. Numerical studies have supported this 
idea\cite{D42}. Our observation in the previous 
section may be a simple realization of the above idea by 
successively applying our analysis to get the 
$D=4 \rightarrow D=3 \rightarrow D=2$ dimensional reduction,
though what we found in this paper is not the Parisi-Sourlas 
type but the Parisi-Wu type of dimensional reduction\footnote{
Recently, Kalkkin and Niemi\cite{KN} have discussed the 
Parisi-Sourlas dimensional reduction in the instanton 
approximation. Connections with our results are unclear.}.
It would be of great interest to investigate low energy dynamics 
of nonabelian gauge theories in a stochastic quantization point
of view furthermore. 

\newpage
%%%%%%%%%%%%%%%%%%%%%%%%%%%%%%%%%%%%%%%%%%%%%%%%%%%%%%%%%%%%%%%

\end{document}